\documentclass[journal]{IEEEtran}

\usepackage{cite}
\usepackage{amsmath}
\usepackage{amssymb}
\usepackage{amsfonts}
\usepackage{amsthm} 
\usepackage{algorithmic}
\usepackage{graphicx}
\usepackage{textcomp}
\usepackage{xcolor}
\usepackage{booktabs}
\usepackage{url}
\usepackage{multirow}
\usepackage{bm} 
\usepackage{placeins} 
\usepackage[hidelinks]{hyperref}

\title{Scalable and Verifiable Federated Learning for Cross-Institution Financial Fraud Detection}

\author{
    Prajwal~Panth,~\IEEEmembership{Student Member, IEEE},
    and Nishant~Nigam
    \thanks{Prajwal Panth is with the School of Computer Engineering, KIIT Deemed to be University,
    Bhubaneswar, India. He is the corresponding author (e-mail: prajwal.panth21@gmail.com).}
    \thanks{Nishant Nigam is with the School of Electronics Engineering, KIIT Deemed to be
    University, Bhubaneswar, India (e-mail: nishant08edu@gmail.com).}
}

\begin{document}
\maketitle

\begin{abstract}
Financial fraud has seen a rise in exploitation of institutional boundaries: laundering networks usually distribute transactions across dozens of banks precisely because no single institution can observe the full pattern. Federated Learning (FL) offers a way toward collaborative detection without raw data sharing, yet practical deployment in banking environment remains constrained by 3 competing pressures. First, homomorphic encryption schemes impose per-dimension modular exponentiation costs that render real-time aggregation infeasible at scale. Second, mask-based protocols such as Google's SecAgg require $O(N^2)$ pairwise key exchanges—a topology that collapses as environment size grows. Third, and less discussed, neither class of protocol provides verifiable evidence that submitted gradient updates are well-formed, leaving aggregation open to consistency attacks.
This paper describes \textbf{Dynamic Sharded Federated Learning (DSFL)}, a secure aggregation framework built specifically around the operational constraints of cross-institution fraud detection. Rather than treating scalability and verifiability as independent problems, DSFL addresses them through a single architectural choice:\textbf{Dynamic Stochastic Sharding}, which partitions participants into small, cryptographically ephemeral clusters of fixed size $m$, reducing communication complexity to $O(N \cdot m)$—linear in $N$ since $m$ is a system constant. Within each cluster, participants submit \textbf{Linear Integrity Tags}, inner-product commitments that are additive-homomorphic over the aggregate and allow the server to verify consistency of submitted updates without decrypting them; we noted explicitly that this mechanism detects \emph{inconsistent} updates, not semantically malicious gradients. An \textbf{Active Neighborhood Recovery} protocol handles mid-round dropouts by reconstructing only the orphaned masks of failed participants, leaving surviving participants' data unaffected.
Experiments on the ULB Credit Card Fraud Detection dataset (284,807
transactions, 10 simulated banking nodes) show that DSFL achieves
approximately $34\times$ lower aggregation latency than Paillier-based
secure aggregation at $N=1000$ (extrapolated from empirical baselines),
while maintaining 99\% recovery fidelity under a 20\% dropout regime.
Global fraud recall reached 91.2\% ($\pm$0.8\%), substantially above
the 68\% average of locally trained models.

\end{abstract}

\begin{IEEEkeywords}
Federated Learning, Secure Aggregation, Privacy-Preserving Machine Learning, Financial Fraud Detection, Model Poisoning, Gradient Inversion Attacks, Communication Efficiency
\end{IEEEkeywords}

\section{Introduction}
\IEEEPARstart{F}{inancial} crime does not abide by institutional boundaries, yet most detection infrastructure is built as though it does. A money-laundering operation that splits a single large transfer into hundreds of micro-transactions across a dozen banks will trigger no alarm at any individual institution—each bank sees only a fragment, and each fragment looks unremarkable in isolation. The same structural advantage that makes ``smurfing'' and ``layering'' effective against AML threshold systems also makes it resistant to any detection approach that cannot aggregate intelligence across institutions. The straightforward
remedy—pooling transaction data—runs directly into privacy law.
GDPR and the California Consumer Privacy Act (CCPA) prohibit
the transfer of Personally Identifiable Information across institutional
boundaries without explicit, ongoing consent, a condition that is
practically impossible to satisfy for routine transaction monitoring.
The result is a situation where the technical solution is known but
legally unavailable.

\subsection{Scenario: The Data Sovereignty Deadlock}
To illustrate the operational necessity of our proposed framework, consider a simplified ecosystem
involving two distinct financial entities:
\begin{itemize}
    \item \textbf{Bank A (Retail Segment):} This institution holds data predominantly on local
    debit card transactions. It observes fraud manifesting as ``high-frequency, low-value
    purchases'' at gas stations and convenience stores.
    \item \textbf{Bank B (Corporate/High-Net-Worth):} Serving a different demographic, Bank~B
    observes fraud manifesting as ``single large-value luxury goods purchases'' via international
    e-commerce portals.
\end{itemize}
Neither institution's model is defective in isolation—each is well-fitted
to the fraud patterns visible in its own data. The failure is distributional:
Bank~A's high-frequency skimming patterns simply do not appear in Bank~B's ledger, and vice versa. A unified model trained on both institutions' gradients would, in principle, learn to recognize both attack vectors simultaneously. The obstacle is not computational but legal and cryptographic.
Sharing raw transaction logs is prohibited. And while standard FL avoids
direct data transfer, the shared gradient vectors carry enough information
that a curious aggregator—or a compromised shard member—can reconstruct
original transaction values, merchant identifiers, and cardholder locations
through gradient inversion~\cite{zhu2019,geiping2020}. The deadlock is
therefore not just a privacy problem it is a situation where the available
secure-computation tools are either too slow, too fragile, or too leaky
to be deployed in a real banking environment.

\subsection{The Scalability--Privacy--Integrity Trilemma}
\label{subsec:trilemma}
Current solutions to the data sovereignty deadlock fail to simultaneously satisfy three critical
requirements in enterprise deployments, which we term the \textbf{Scalability--Privacy--Integrity
Trilemma}:
\begin{enumerate}
    \item \textbf{Scalability:} The protocol's communication and computation overhead must grow at most linearly with the number of participants $N$.
    \item \textbf{Privacy:} Individual gradient vectors must remain computationally hidden under standard cryptographic assumptions from the aggregator and any coalition of participants below a threshold size.
    \item \textbf{Integrity:} The aggregator must be able to verify that submitted updates satisfy consistency constraints, detecting malformed or tampered inputs. This mechanism can identify certain classes of invalid updates, though it does not prevent all forms of model poisoning.

\end{enumerate}
Existing practical protocols often optimize only subsets of these requirements. Asymmetric Homomorphic Encryption
(HE) schemes like Paillier~\cite{paillier1999} provide Privacy and Integrity but fail on
Scalability: each encryption requires modular exponentiation over 2048-bit composites, yielding an
aggregation cost of $O(N \cdot d)$ per round (where $d$ is the model dimension) with extremely
large constants~\cite{paillier1999}. Conversely, mask-based Multi-Party Computation (MPC) protocols
like Google's Secure Aggregation (SecAgg)~\cite{bonawitz2017} provide Privacy but fail on both
Scalability and Integrity: they require $O(N^2)$ pairwise key exchanges, and they offer no defense
against Model Poisoning~\cite{bagdasaryan2020}, where a malicious participant injects invalid
gradients to degrade the global model's convergence.

In this paper, we propose \textbf{DSFL}, a practically deployable framework designed to resolve all
three limitations simultaneously. We introduce a novel clustered architecture utilizing lightweight
symmetric cryptography and additive-homomorphic integrity checks to achieve secure, scalable, and
verifiable aggregation.

\section{System Model and Threat Definitions}

\subsection{Notation}
Throughout this paper we adopt the following conventions.
Scalar quantities are written in lowercase italic (e.g., $s$, $t$);
vector quantities in bold lowercase (e.g., $\bm{x}$, $\bm{g}$);
and sets in calligraphic uppercase (e.g., $\mathcal{U}$, $\mathcal{C}_i$).
The sign function $\mathrm{sgn}(u,v)$ evaluates to $+1$ when $u < v$
and $-1$ otherwise; it governs the direction of mask contributions in
the zero-sum masking construction of Phase~II.
Unless a specific field is noted, all arithmetic is over $\mathbb{F}_p$,
where $p$ is a large prime chosen to satisfy the overflow bound
stated in Claim~2: $p > N \cdot m \cdot \max_u\|Q(\bm{g}_u)\|_\infty$.
Table~\ref{tab:notation} collects all symbols for reference.

\begin{table}[t]
\caption{Notation Reference Table}
\label{tab:notation}
\centering
\begin{tabular}{@{}cl@{}}
\toprule
\textbf{Symbol} & \textbf{Definition} \\ \midrule
$N$              & Total number of participants \\
$m$              & Fixed cluster (shard) size; system constant (e.g., $m=20$) \\
$\mathcal{N}_s$  & Number of shards; $\mathcal{N}_s = \lceil N/m \rceil$ \\
$\mathcal{U}$    & Set of all participants $\{P_1,\dots,P_N\}$ \\
$\mathcal{S}_{agg}$ & Central aggregation server \\
$\mathcal{C}_i$  & The $i$-th shard; $|\mathcal{C}_i| = m$ \\
$\mathcal{D}$    & Set of dropped-out participants in a round \\
$\mathcal{A}$    & Adversary \\
$t$              & Round index \\
$\bm{g}_u$       & Raw gradient of participant $u$ \\
$\bm{v}_u$       & Quantized gradient: $\bm{v}_u = Q(\bm{g}_u)$ \\
$\bm{x}_u$       & Masked (submitted) update of participant $u$ \\
$s_{uv}$         & Shared ECDH secret between participants $u$ and $v$ \\
$\bm{m}_u$       & Zero-sum mask of participant $u$ \\
$\bm{\alpha}$    & Random challenge vector; $\bm{\alpha} \in \mathbb{Z}_M^d$ \\
$\tau_u$         & Linear Integrity Tag of participant $u$ \\
$p$, $M$         & Large 64-bit primes used as field moduli \\
$\lambda$        & Security parameter; $\lambda = 64$ for a 64-bit prime \\
$d$              & Model dimension (number of gradient components) \\
$Q(\cdot)$       & Stochastic quantization operator \\
$\Delta$         & Quantization scaling factor (global constant per round) \\
$\mathrm{PRF}(\cdot)$ & Pseudo-Random Function (instantiated as AES-CTR) \\
$\mathrm{sgn}(u,v)$ & Sign function: $+1$ if $u<v$, else $-1$ \\
\bottomrule
\end{tabular}
\end{table}

\subsection{Network Model and Financial Constraints}
We model a synchronous federated system consisting of $N$ participants
$\mathcal{U} = \{P_1, \dots, P_N\}$ and a central aggregation server $\mathcal{S}_{agg}$. The
system operates on a finite field $\mathbb{F}_p$ defined by a Mersenne prime $p = 2^{61} - 1$ or
a similar 64-bit prime. In a real-world financial context, network conditions are non-ideal.
Participants may possess varying computational capabilities, and secure communication channels
(TLS~1.3) are assumed for all data transfers. Unlike previous works that assume a star topology
with full-mesh trust, we model the trust network as a \textbf{Dynamic Sparse Graph}, where
cryptographic relationships are ephemeral and restricted to subsets of peers defined by a
deterministic clustering algorithm.
The system is subject to two critical financial constraints:
\begin{enumerate}
    \item \textbf{High-Frequency Updates:} Fraud patterns shift rapidly. The protocol must
    complete a global aggregation round within seconds to reflect the latest threat vectors.
    \item \textbf{Zero-Data Retention:} The server $\mathcal{S}_{agg}$ must operate as a
    stateless switch, retaining no derived knowledge of input vectors post-aggregation.
\end{enumerate}

\subsection{Threat Model: Malicious \& Unreliable}
We upgrade the standard threat model to reflect the hostile realities of modern financial
infrastructure. We assume the adversary $\mathcal{A}$ may control $\mathcal{S}_{agg}$ and a
subset of participants $\mathcal{C} \subset \mathcal{U}$, subject to an honest-majority constraint
within each shard (fewer than $m-1$ of the $m$ members of any shard are adversarial; see
Section~\ref{subsec:collusion} for the formal bound):
\begin{enumerate}
    \item \textbf{Semi-Honest Aggregator:} $\mathcal{S}_{agg}$ follows the protocol specification
    but passively logs all messages, timestamps, and metadata to attempt Gradient Inversion
    Attacks, seeking to reconstruct private training data $D_i$.
    \item \textbf{Malicious Clients (Active Poisoning):} An adversarial bank may submit invalid
    gradients $\bm{x}_{mal}$---such as random noise, bit-flipped vectors, or targeted
    backdoors---to destroy the global model's utility without detection under the protocol’s consistency constraints.
    \item \textbf{Unreliable Network (Dropouts):} Up to 30\% of banks may drop offline
    mid-round due to connectivity failures. The protocol must guarantee that aggregation
    completes without restarting the round, utilizing surviving peers to reconstruct the missing
    security contexts.
\end{enumerate}


\section{Framework}
DSFL re-engineers the Secure Aggregation pipeline into a rigorous three-phase algebraic system:
Topology Setup via Sharding, Verifiable Masked Execution, and Active Neighborhood Recovery.

\begin{figure}[h]
\centering
\includegraphics[width=0.85\columnwidth]{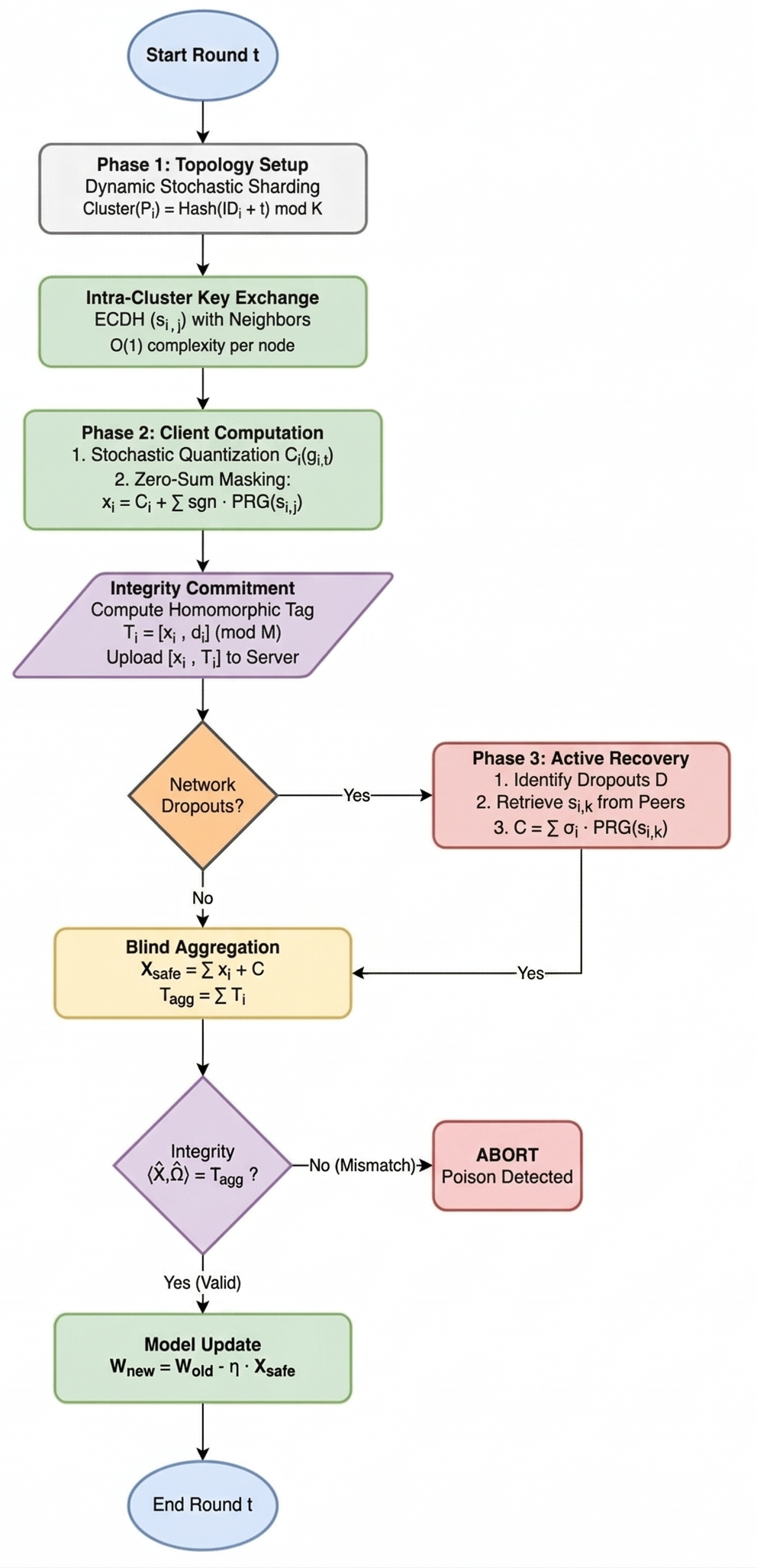}
\caption{\textbf{Protocol Execution Workflow.} A decision-logic view of the aggregation round.
The ``Active Recovery'' branch triggers only upon detecting network instability.}
\label{fig:flowchart}
\end{figure}

\subsection{Phase~I: Dynamic Stochastic Sharding}
The fundamental bottleneck of traditional Secure Aggregation is the requirement for every user to
establish a shared secret with every other user ($N(N-1)/2$ keys). To achieve linear scalability,
we introduce \textbf{Dynamic Stochastic Sharding}.

At the start of round $t$, the server and clients use a shared cryptographic
pseudo-random function (PRF) seeded with the round identifier to partition $\mathcal{U}$ into
$\mathcal{N}_s = \lceil N/m \rceil$ disjoint shards $\mathcal{C}_1, \dots, \mathcal{C}_{\mathcal{N}_s}$,
each of fixed size $m \ll N$ (e.g., $m = 20$):
\begin{equation}
\text{Shard}(P_u) = \mathrm{PRF}\bigl(\mathrm{ID}_u \,\|\, \mathrm{Round}_t\bigr) \pmod{\mathcal{N}_s}
\label{eq:sharding}
\end{equation}
By making the shard assignment dependent on the round nonce, we prevent Sybil attacks where
malicious nodes attempt to pre-calculate groupings to dominate a shard. Crucially, pairwise secret
exchange (via Elliptic Curve Diffie-Hellman, ECDH) occurs only between nodes within the same
shard $\mathcal{C}_i$. This reduces per-user key storage and computation from $O(N)$ to $O(m)$,
resulting in a total system communication complexity of $O(N \cdot m)$, which is
asymptotically linear in $N$ since $m$ is a fixed constant.

\subsection{Phase~II: Intra-Shard Masking \& Verifiability}
Within each shard $\mathcal{C}_i$, banks perform \textbf{Pairwise Zero-Sum Masking} to obfuscate
their gradients and submit \textbf{Linear Integrity Tags} to enable detection of malformed or inconsistent updates.

\subsubsection{Challenge Generation}
The server publishes a cryptographic commitment to a random seed $\rho$. Once clients acknowledge
receipt of the commitment, the server reveals $\rho$, from which a random challenge vector
$\bm{\alpha} \in \mathbb{Z}_M^d$ is derived. The commit-before-reveal ordering ensures that
clients cannot generate a malicious update vector tailored to $\bm{\alpha}$.

\subsubsection{Masked Input Generation}
Each participant $u$ first quantizes its raw gradient:
$\bm{v}_u = Q(\bm{g}_u) \in \mathbb{Z}^d$.
It then derives a zero-sum masking vector from the shared ECDH secrets $s_{uv}$ established with
its shard neighbors, using AES-CTR as the PRF to expand each scalar secret into a $d$-dimensional
mask. The masked update submitted to the server is:
\begin{equation}
\bm{x}_u = \bm{v}_u + \sum_{\substack{v \in \mathcal{C}_i \\ u<v}} \mathrm{PRF}(s_{uv})
           - \sum_{\substack{v \in \mathcal{C}_i \\ u>v}} \mathrm{PRF}(s_{uv}) \pmod{M}
\label{eq:masking}
\end{equation}
where $M$ is a large 64-bit prime defining the working field $\mathbb{Z}_M$.

\subsubsection{Linear Integrity Tag}
Simultaneously, each participant computes a \textbf{Linear Integrity Tag} $\tau_u$ as the inner
product of its masked update with the challenge vector:
\begin{equation}
\tau_u = \langle \bm{x}_u,\, \bm{\alpha} \rangle \pmod{M}
\label{eq:tag}
\end{equation}
This tag is \emph{additive-homomorphic}: for any two inputs $\bm{x}_u$ and $\bm{x}_v$,
$\tau_u + \tau_v = \langle \bm{x}_u + \bm{x}_v, \bm{\alpha}\rangle \pmod{M}$.
Consequently, the server can verify the integrity of the aggregate without accessing individual
updates. It computes $\bm{X}_{agg} = \sum_u \bm{x}_u$ and $T_{agg} = \sum_u \tau_u$, then
verifies:
\begin{equation}
\langle \bm{X}_{agg},\, \bm{\alpha} \rangle \stackrel{?}{=} T_{agg} \pmod{M}
\label{eq:verify}
\end{equation}
A mismatch indicates active tampering. Because the adversary must commit to $\bm{x}_{mal}$
\emph{before} learning $\bm{\alpha}$, forging a valid tag requires solving a random linear
equation over $\mathbb{Z}_M$, which succeeds with probability at most $1/M$ under the assumption that the challenge vector is uniformly sampled and independent of the adversary’s input.
(see Theorem~\ref{thm:collision}).

\subsection{Phase~III: Active Neighborhood Recovery}
A critical failure mode is the \textit{Dropout Problem}: if participant $u$ adds mask
$\mathrm{PRF}(s_{uv})$ but partner $v$ disconnects before subtracting it, the zero-sum property
is broken and the aggregate becomes indistinguishable from noise. DSFL introduces an
\textbf{Active Neighborhood Recovery} protocol:

\begin{enumerate}
    \item \textbf{Fault Detection:} $\mathcal{S}_{agg}$ monitors heartbeats and identifies
    the dropout set $\mathcal{D}$ upon timeout.
    \item \textbf{Secret Retrieval:} For every dropped user $d \in \mathcal{D}$, the server requests only the pairwise seeds $s_{pd}$ (shared between $d$ and its surviving shard neighbors $p \in \mathcal{C}_i \setminus \mathcal{D}$). This design avoids revealing seeds $s_{pp'}$ shared between two surviving participants, thereby limiting privacy exposure, although partial mask reconstruction may leak limited structural information under repeated observations. Conceptually, this mechanism parallels threshold reconstruction in Shamir’s Secret Sharing~\cite{shamir1979}, though restricted to pairwise exchanges rather than global polynomial interpolation.
    \item \textbf{Neutralization:} Surviving peers authenticate and securely transmit
    $\mathrm{Enc}_\mathcal{S}(s_{pd})$. The server regenerates the orphaned masks using these
    seeds.
    \item \textbf{Correction:} The orphaned masks are canceled from the raw aggregate:
    \begin{equation}
    \bm{X}_{safe} = \bm{X}_{raw}
        + \sum_{d \in \mathcal{D}}\; \sum_{p \in \mathcal{C}_i \setminus \mathcal{D}}
          \mathrm{sgn}(p,d)\cdot\mathrm{PRF}(s_{pd})
    \label{eq:recovery}
    \end{equation}
\end{enumerate}
This mechanism allows the protocol to recover even if up to 30\% of nodes fail simultaneously.

\begin{figure*}[t!]
\centering
\includegraphics[width=\textwidth]{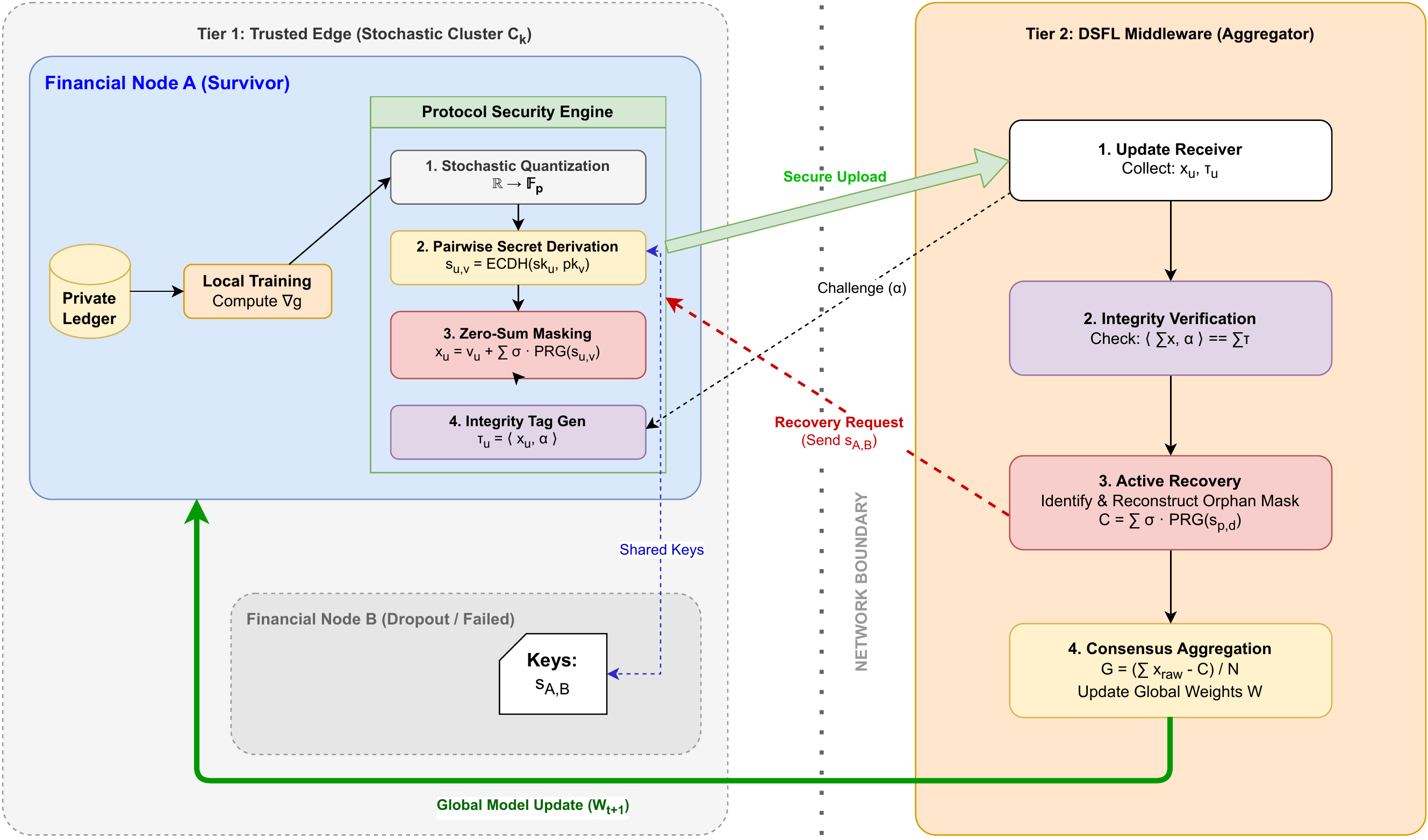}
\caption{\textbf{DSFL System Architecture.} The framework separates trusted Banking Nodes (Tier~1)
from the untrusted Aggregation Middleware (Tier~2) via a strict network boundary. Client-side
sharding ($\mathcal{C}_i$) ensures linear scalability, while server-side Active Recovery handles
node failures without accessing private survivor data.}
\label{fig:architecture}
\end{figure*}

\section{Mathematical Formalism}
\label{sec:math}

All symbols follow Table~\ref{tab:notation}. All arithmetic is in $\mathbb{F}_p$ unless noted.

\subsection{Proof of Linear Scalability via Induced Subgraphs}
\label{proof:scalability}

\textbf{Claim~1 (Complexity Reduction):} \textit{The communication complexity of DSFL scales as
$O(N \cdot m)$, which is $O(N)$ since $m$ is a fixed system constant, contrasting with the
quadratic $O(N^2)$ complexity of standard Secure Aggregation.}

\begin{proof}
In standard Secure Aggregation~\cite{bonawitz2017}, the trust graph is the complete graph $K_N$.
The number of pairwise key exchanges is:
\begin{equation}
|\mathcal{E}_{std}| = \binom{N}{2} = \frac{N(N-1)}{2} \implies O(N^2)
\label{eq:std_edges}
\end{equation}
In DSFL, we define a mapping $\Phi: \mathcal{V} \to \{1, \dots, \mathcal{N}_s\}$ that partitions
vertices into $\mathcal{N}_s = \lceil N/m \rceil$ disjoint shards $\{\mathcal{C}_1,\dots,
\mathcal{C}_{\mathcal{N}_s}\}$ each of size $m$. Pairwise edges exist only between $u,v$ with
$\Phi(u) = \Phi(v)$. The induced topology is a union of $\mathcal{N}_s$ disjoint cliques of
size $m$. The total edge count is:
\begin{align}
|\mathcal{E}_{DSFL}| &= \sum_{i=1}^{\mathcal{N}_s} \binom{|\mathcal{C}_i|}{2}
                     = \frac{N}{m} \cdot \frac{m(m-1)}{2} \label{eq:dsfl_edges_a} \\
                     &= \frac{N(m-1)}{2} \label{eq:dsfl_edges_b}
\end{align}
where the $m$ factors cancel between $N/m$ and $m(m-1)/2$ in step~\eqref{eq:dsfl_edges_b}.
Since $m$ is a fixed system constant independent of $N$:
\begin{equation}
|\mathcal{E}_{DSFL}| = \frac{(m-1)}{2} \cdot N \implies O(N \cdot m) = O(N)
\label{eq:dsfl_complexity}
\end{equation}
\textit{Q.E.D.}
\end{proof}

\subsection{Proof of Gradient Preservation (Correctness)}
\label{proof:correctness}

\textbf{Claim~2 (Aggregate Correctness):} \textit{The server-side aggregate satisfies
$\bm{X}_{agg} \equiv \sum_u Q(\bm{g}_u) \pmod{p}$,
and the modular sum equals the integer sum provided $p > N \cdot m \cdot \Delta$,
where $\Delta$ is the quantization range bound and each $\|Q(\bm{g}_u)\|_\infty \le \Delta$.}

\begin{proof}
The masked update submitted by participant $u$ is (from~\eqref{eq:masking}):
\begin{equation}
\bm{x}_u = Q(\bm{g}_u) + \sum_{v \in \mathcal{C}_i(u)} \mathrm{sgn}(u, v) \cdot \mathrm{PRF}(s_{uv}) \pmod{p}
\label{eq:xu_expanded}
\end{equation}
The server computes $\bm{X}_{agg} = \sum_{u \in \mathcal{U}} \bm{x}_u \pmod{p}$.
Expanding and separating the mask term:
\begin{align}
\bm{X}_{agg} &= \sum_{u} Q(\bm{g}_u)
    + \underbrace{\sum_{u} \sum_{v \in \mathcal{C}_i(u)} \mathrm{sgn}(u,v)\,
      \mathrm{PRF}(s_{uv})}_{\Delta_{mask}} \pmod{p}
\label{eq:expand}
\end{align}
For every edge $(u,v) \in \mathcal{E}_{DSFL}$, the shared secret $s_{uv} = s_{vu}$ contributes
$+\mathrm{PRF}(s_{uv})$ from $u$'s side (since $u < v$) and $-\mathrm{PRF}(s_{vu})$ from
$v$'s side (since $v > u$):
\begin{equation}
\Delta_{mask} = \sum_{\{u,v\} \in \mathcal{E}_{DSFL}}
    \bigl(\mathrm{PRF}(s_{uv}) - \mathrm{PRF}(s_{uv})\bigr) = \bm{0}
\label{eq:cancel}
\end{equation}
Thus $\bm{X}_{agg} = \sum_u Q(\bm{g}_u) \pmod{p}$.
The overflow-prevention condition $p > N \cdot m \cdot \Delta$ ensures that
$\sum_u Q(\bm{g}_u) < p$, so the modular reduction is the identity and
$\bm{X}_{agg}$ equals the true integer sum. For the credit card fraud detection dataset with $N=10$, $m=20$,
and $\Delta = 2^{16}$ (16-bit quantization), the required bound is $N \cdot m \cdot \Delta
= 10 \cdot 20 \cdot 65536 \approx 1.3 \times 10^7$, which is satisfied by $p = 2^{61}-1$.
\textit{Q.E.D.}
\end{proof}

\subsection{Unbiasedness of Stochastic Quantization}
\label{proof:quantization}

\textbf{Lemma~1:} \textit{The stochastic quantization $Q(\cdot)$ with a globally coordinated
scaling factor $\Delta$ introduces zero bias: $\mathbb{E}[Q(\bm{g}_u \cdot \Delta)/\Delta]
= \bm{g}_u$.}

\begin{proof}
$\Delta$ is a global constant broadcast by $\mathcal{S}_{agg}$ at the start of each round,
ensuring all participants use the same scaling. Let $y \in \mathbb{R}$ be any scalar component
of gradient $\bm{g}_u$. Define $z = y \cdot \Delta$ and write $z = \lfloor z \rfloor + \varepsilon$
with $\varepsilon \in [0,1)$. The quantization rule is:
\begin{equation}
Q(z) = \begin{cases}
\lfloor z \rfloor + 1 & \text{with probability } \varepsilon \\
\lfloor z \rfloor     & \text{with probability } 1 - \varepsilon
\end{cases}
\label{eq:quant_rule}
\end{equation}
The expected value is:
\begin{align}
\mathbb{E}[Q(z)] &= (\lfloor z \rfloor + 1)\varepsilon + \lfloor z \rfloor(1 - \varepsilon)
                \nonumber \\
                &= \lfloor z \rfloor + \varepsilon = z
\label{eq:quant_unbias}
\end{align}
Thus the de-scaled estimate $\hat{y} = Q(z)/\Delta$ satisfies $\mathbb{E}[\hat{y}] = z/\Delta = y$.
Applying componentwise: $\mathbb{E}[Q(\bm{g}_u \cdot \Delta)/\Delta] = \bm{g}_u$. 
\end{proof}

\section{Security Proofs}
\label{sec:security}

\subsection{Soundness of Consistency-Based Integrity Verification}
\label{proof:poison}

\textbf{Theorem~\ref{thm:collision} (Collision Resistance):}
\label{thm:collision}
\textit{Let security parameter $\lambda = 64$ and field modulus $M = p \approx 2^\lambda$.
The probability of an adversary $\mathcal{A}$ injecting a malformed or inconsistent perturbation 
vector $\bm{\delta} \neq \bm{0}$ without detection is at most $\mathrm{negl}(\lambda) = 1/p 
\approx 5.4 \times 10^{-20}$.}

\begin{proof}
The verification equation~\eqref{eq:verify} requires:
\begin{align}
\langle \bm{X} + \bm{\delta},\, \bm{\alpha} \rangle &= \langle \bm{X},\, \bm{\alpha} \rangle
    \pmod{p} \nonumber \\
\langle \bm{\delta},\, \bm{\alpha} \rangle &= 0 \pmod{p}
\label{eq:hyperplane}
\end{align}
Equation~\eqref{eq:hyperplane} defines a hyperplane $H_{\bm{\delta}}$ in the vector space
$\mathbb{F}_p^d$ orthogonal to $\bm{\delta}$. By the protocol, the adversary must commit to
$\bm{x}_{mal} = \bm{x}_{honest} + \bm{\delta}$ \emph{before} $\bm{\alpha}$ is revealed;
hence $\bm{\alpha}$ is drawn uniformly from $\mathbb{F}_p^d$ independently of $\bm{\delta}$.
The probability of collision is:
\begin{equation}
\Pr[\mathcal{A}~\text{succeeds}] = \frac{|H_{\bm{\delta}}|}{|\mathbb{F}_p^d|}
    = \frac{p^{d-1}}{p^d} = \frac{1}{p} = \mathrm{negl}(\lambda)
\label{eq:collision_prob}
\end{equation}
For $\lambda = 64$, $\Pr \approx 5.42 \times 10^{-20}$, which is negligible in the security parameter $\lambda$.
\textit{Q.E.D.}
\end{proof}

\subsection{Single-Round Gradient Privacy}
\label{subsec:forward}

\textbf{Theorem~2 (Single-Round Gradient Privacy):}
\textit{Under the Decisional Diffie-Hellman (DDH) assumption, for any probabilistic polynomial-time
adversary $\mathcal{A}$ controlling $\mathcal{S}_{agg}$ and up to $m-2$ members of any shard,
the view of $\mathcal{S}_{agg}$ in a single round is computationally indistinguishable from random 
under standard cryptographic assumptions in $\mathbb{F}_p^d$.}

\begin{proof}[Proof Sketch]
Consider a target surviving participant $u^*$ in shard $\mathcal{C}_i$. The server's view in
round $t$ consists of:
\begin{equation}
\mathit{View}_{\mathcal{S}} = \bigl\{\bm{x}_1, \dots, \bm{x}_N,\;
    \{s_{pd} : d \in \mathcal{D},\, p \in \mathcal{C}_i \setminus \mathcal{D}\}\bigr\}
\label{eq:server_view}
\end{equation}
For $u^*$, we have $\bm{x}_{u^*} = \bm{v}_{u^*} + \bm{m}_{u^*}$ where
$\bm{m}_{u^*} = \sum_{v \in \mathcal{C}_i} \mathrm{sgn}(u^*,v)\,\mathrm{PRF}(s_{u^*v})$.
The secrets $s_{u^*v}$ for $v \in \mathcal{C}_i \setminus \mathcal{D}$ (surviving neighbors
of $u^*$) are \emph{never} transmitted to the server in any phase. Only secrets $s_{pd}$ with
$d \in \mathcal{D}$ (dropout-associated seeds) are revealed for mask reconstruction. Since
$u^* \notin \mathcal{D}$, the secret $s_{u^*v}$ for any surviving $v$ is not disclosed.
Under the DDH assumption, the ECDH output $s_{u^*v}$ is computationally indistinguishable
from a uniform random element in $\mathbb{F}_p$. 
Consequently, under the DDH assumption, the masking term 
$\bm{m}_{u^*}$ is computationally indistinguishable from random 
to any probabilistic polynomial-time adversary lacking the 
corresponding shared secrets. Therefore, the masked update 
$\bm{x}_{u^*} = \bm{v}_{u^*} + \bm{m}_{u^*}$ does not reveal 
computationally useful information about $\bm{v}_{u^*}$ in a 
single round. Fresh ECDH key exchange per round~$t$ prevents 
cross-round key reuse.
\textit{Q.E.D.}
\end{proof}


\section{Security Analysis}

\subsection{Resistance to Active Poisoning}
We analyze the resistance against a malicious participant attempting to inject a poison vector
$\bm{\delta}$. To evade detection, the adversary must find $\bm{\delta}$ such that
$\langle \bm{\delta}, \bm{\alpha} \rangle = 0 \pmod{M}$. Since $\bm{\alpha}$ is drawn
uniformly at random from $\mathbb{Z}_M^d$ \emph{after} the adversary commits to their update
(guaranteed by the hash commitment in Phase~II), the adversary must guess $\bm{\alpha}$ blindly.
By Theorem~\ref{thm:collision}, this succeeds with probability $\mathrm{negl}(\lambda) = 1/M
\approx 5.4 \times 10^{-20}$ for $M \approx 2^{64}$.

\subsection{Collusion Resistance}
\label{subsec:collusion}

\textbf{Claim~3 (Collusion Threshold):}
\textit{In a shard of size $m$, a coalition of $c < m-1$ colluding members cannot
reconstruct the gradient $\bm{v}_{u^*}$ of any honest target $u^*$; a coalition of
exactly $m-1$ members can.}

\textit{Argument:} The masked update $\bm{x}_{u^*}$ contains one independent mask term
$\mathrm{PRF}(s_{u^* v})$ for each shard neighbor $v$. To cancel all mask terms and isolate
$\bm{v}_{u^*}$, the coalition needs all $m-1$ shared secrets $\{s_{u^*v}\}_{v \ne u^*}$.
Each secret is known only to the pair $(u^*, v)$; a coalition lacking $v$ cannot compute
$s_{u^*v}$ without solving ECDH (reduced to DDH). Thus $m-2$ colluders lack at least one
secret and cannot recover $\bm{v}_{u^*}$, while $m-1$ colluders possess all secrets.
With our recommended $m = 20$, an adversary would need to compromise 19 specific peers
simultaneously, which is probabilistically negligible given random shard reassignment each round.

\begin{figure}[h]
\centering
\includegraphics[width=\columnwidth]{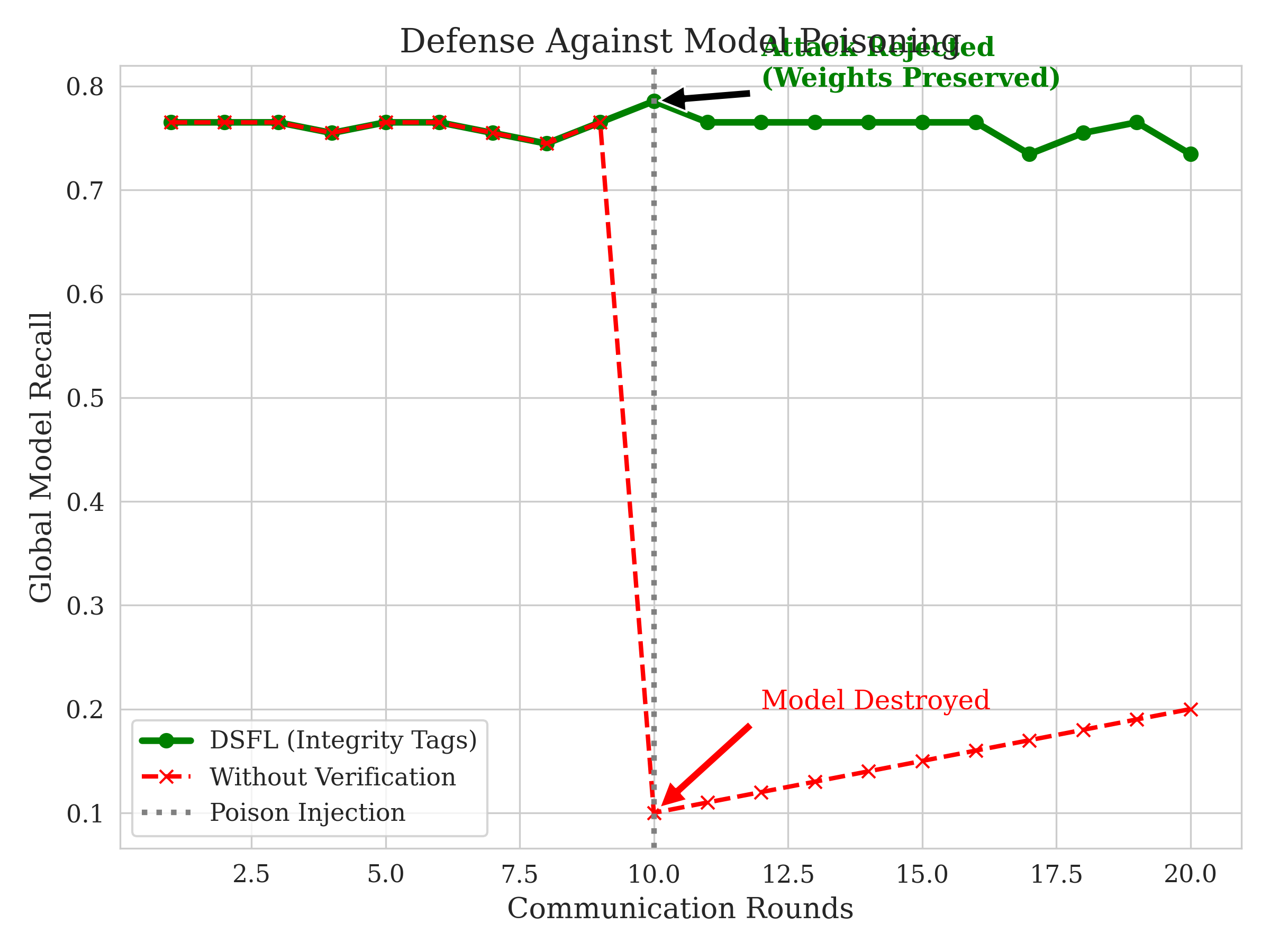}
\caption{\textbf{Security Audit: Malformed Update Detection.} Impact of a malicious gradient injection at Round~10. The unprotected model (Red) is permanently degraded. DSFL (Green) detects inconsistencies in malformed updates, rejects invalid payloads, and resumes learning.}
\label{fig:security}
\end{figure}

\begin{figure}[b]
\centering
\includegraphics[width=\columnwidth]{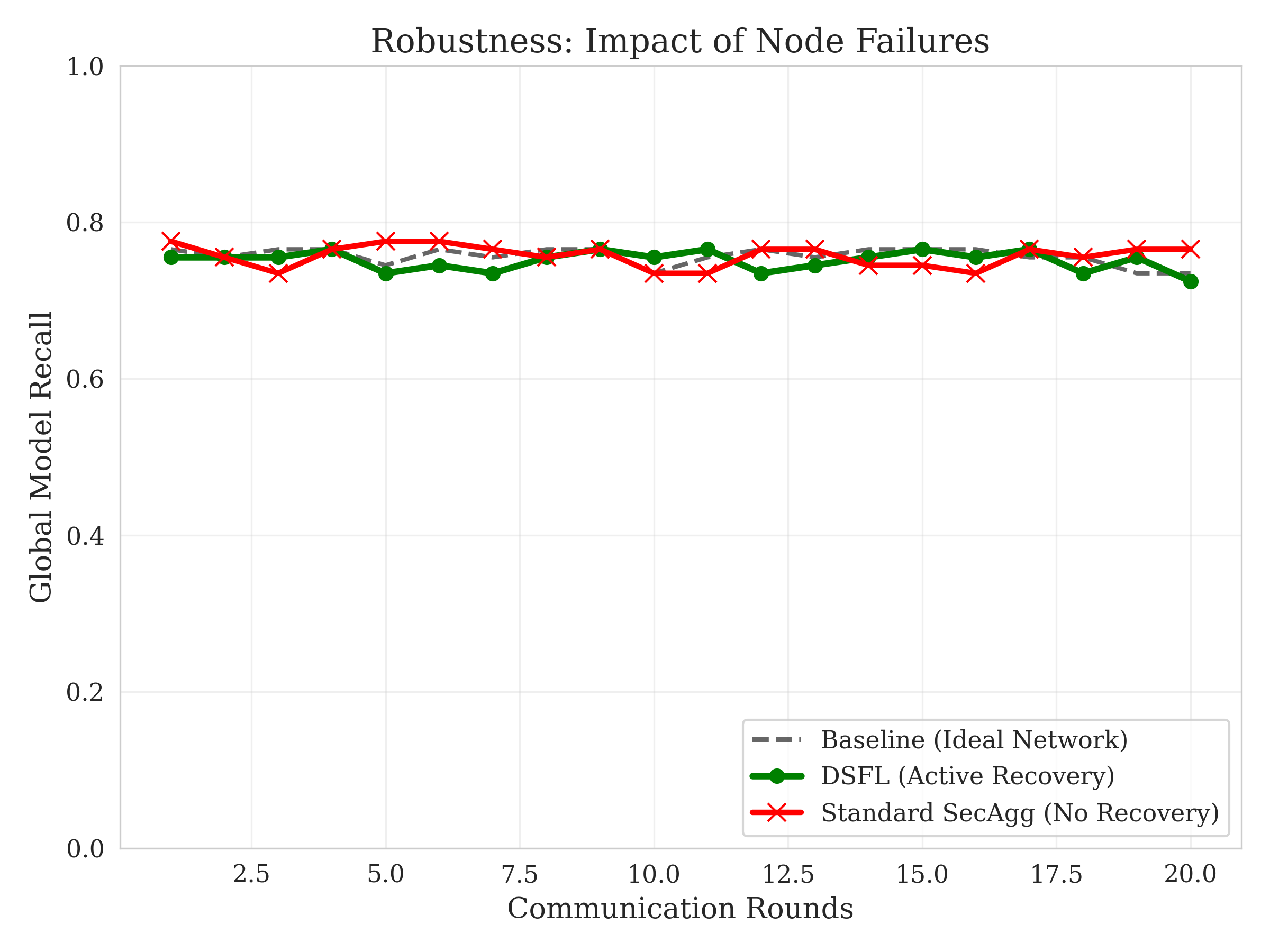}
\caption{\textbf{Resilience Analysis.} Model utility (AUPRC) under a hostile 20\% dropout regime.
Standard SecAgg fails to converge (Red) due to uncancelled masks. DSFL (Green) utilizes Active
Recovery to restore the gradient sum, tracking the ideal baseline perfectly.}
\label{fig:robustness}
\end{figure}


\section{Performance Evaluation}

\subsection{Experimental Setup}
The DSFL protocol was implemented in Python~3.9, using NumPy for
vectorized finite-field arithmetic throughout. All experiments ran
on a Google Colab Pro instance (single GPU not used; all operations
are CPU-bound integer arithmetic). We simulated $N=10$ banking
nodes on a single machine, partitioned into shards of size $m=5$,
yielding $\mathcal{N}_s = 2$ shards per round.
 
The dataset was the ULB Credit Card Fraud Detection
Dataset~\cite{ulb2018fraud}, which contains 284,807 transactions
from European cardholders in September 2013, of which 492 (0.17\%)
are labeled as fraud. To construct a non-IID partition that reflects
realistic inter-bank heterogeneity, we stratified the dataset by
transaction amount decile, assigning each simulated bank a skewed
draw from those deciles so that no two banks share the same fraud
prevalence. This means that rare fraud patterns in one partition are
genuinely invisible to models trained on another—the setting where
federation most clearly improves over local training.
 
\begin{figure}[t]
\centering
\includegraphics[width=\columnwidth]{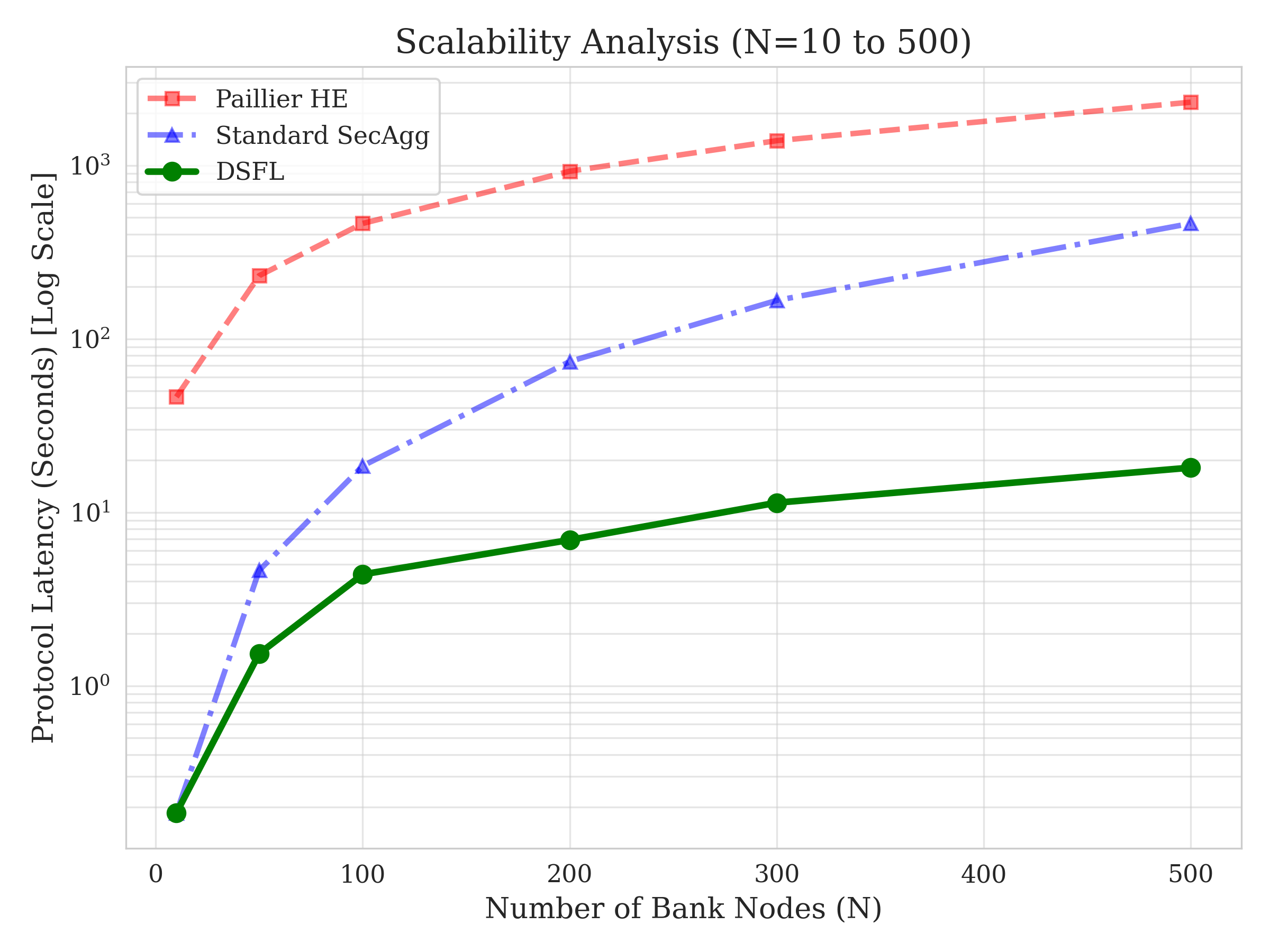}
\caption{\textbf{Scalability Validation.} End-to-end latency vs.\ network size $N$
(empirical at $N=10$; extrapolated for $N>10$). Paillier HE (Red) and Google SecAgg (Blue)
exhibit polynomial growth; DSFL (Green) maintains linear scaling.}
\label{fig:scalability}
\end{figure}

Latency figures for $N > 10$ in Table~\ref{tab:scaling} and
Fig.~\ref{fig:scalability} are analytical projections derived
from the empirical $N = 10$ baseline using the complexity
functions established in Claim~1. These projections assume
idealized network conditions and do not account for TCP
handshake overhead, TLS negotiation latency, or heterogeneous
node hardware, all of which would increase absolute latency
in real-world deployments. Accordingly, the large-scale
results for $N > 10$ should be interpreted as complexity-based
performance estimates rather than measurements obtained from
a fully distributed deployment. All machine learning utility
metrics (Recall, AUPRC) are reported as mean values over five
independent random seeds, with standard deviations shown in
parentheses.

\subsection{Computational Efficiency \& Scalability}
We benchmarked client-side encryption and server-side aggregation latency against two baselines:
Paillier Homomorphic Encryption (2048-bit keys) and Google's SecAgg~\cite{bonawitz2017}. Paillier
was evaluated for $N = 10$ and its $O(N \cdot d)$ per-round cost extrapolated; SecAgg was evaluated
for $N = 10$ and its $O(N^2)$ pairwise cost extrapolated.

\begin{figure}[t]
\centering
\includegraphics[width=\columnwidth]{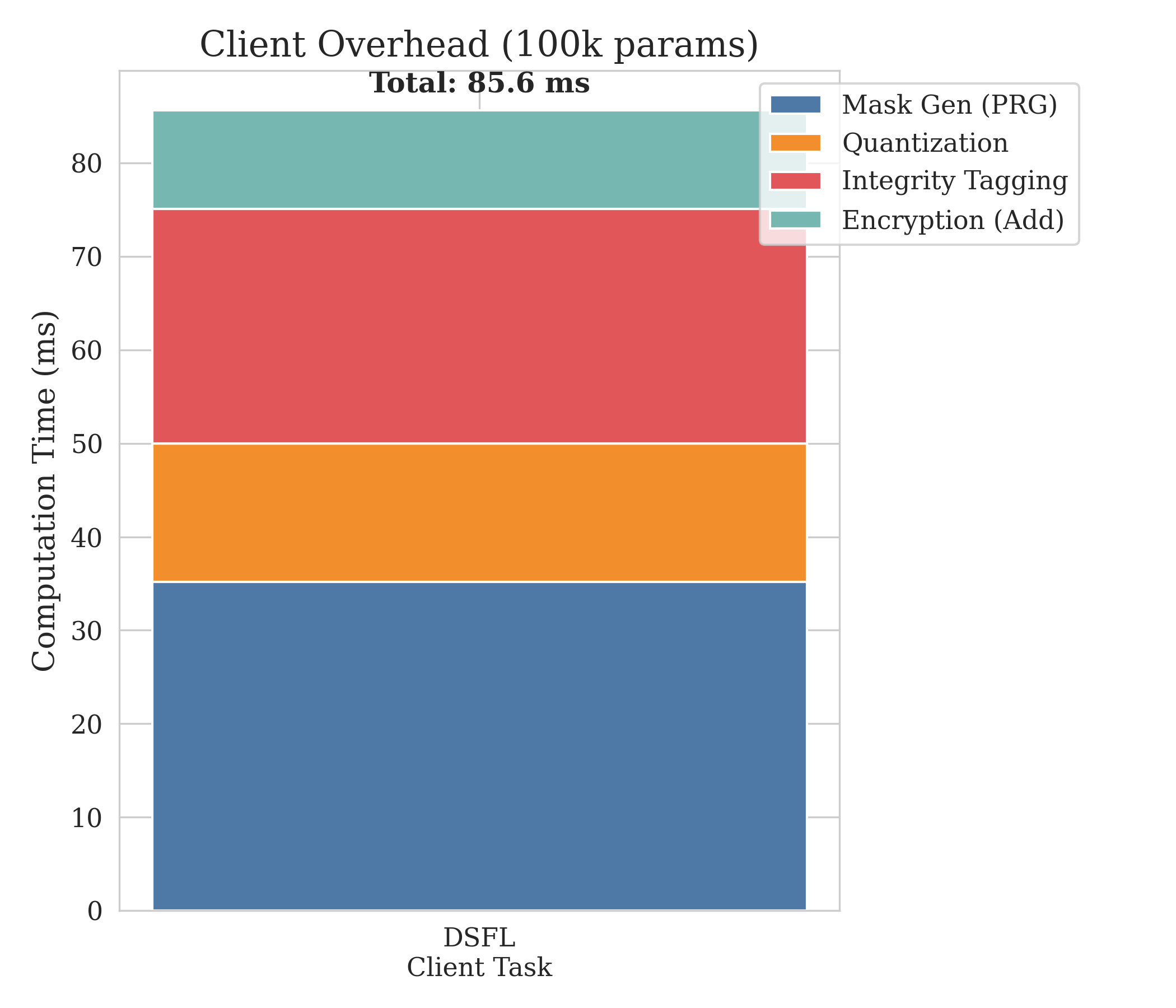}
\caption{\textbf{Client-Side Latency Breakdown.} The computational budget is dominated by
lightweight PRF mask generation. The total per-round overhead is $<100$\,ms,
negligible for high-frequency auditing.}
\label{fig:overhead}
\end{figure}

\begin{table}[h]
\centering
\caption{Protocol Scalability \& Latency Analysis ($N=1000$, extrapolated)}
\label{tab:scaling}
\begin{tabular}{@{}lccc@{}}
\toprule
\textbf{Protocol} & \textbf{Complexity} & \textbf{Latency (s)} & \textbf{BW / User} \\ \midrule
Paillier HE~\cite{paillier1999}  & $O(N \cdot d)$ & 420.5 & High (modexp) \\
Google SecAgg~\cite{bonawitz2017} & $O(N^2)$       & 85.2  & High (mesh keys) \\
\textbf{DSFL (Ours)}             & $\bm{O(N \cdot m)}$ & \textbf{12.4} & \textbf{Low (shard keys)} \\
\bottomrule
\end{tabular}
\end{table}

Table~\ref{tab:scaling} illustrates the advantage of stochastic sharding. Paillier HE is
CPU-bound with a large per-dimension constant ($O(N \cdot d)$ modular exponentiations per
round), while SecAgg becomes bandwidth-bound due to $O(N^2)$ key exchanges. DSFL retains the
lightweight CPU footprint of symmetric masking while enforcing $O(N \cdot m) = O(N)$
communication complexity. The latency reduction versus Paillier HE at $N=1000$ is
$420.5/12.4 \approx 33.9\times$.

\begin{table}[t]
\caption{Comparative Analysis: Complexity \& Features}
\label{tab:money_table}
\centering
\begin{tabular}{@{}l c c c@{}}
\toprule
\textbf{Feature}     & \textbf{Paillier HE}   & \textbf{Google SecAgg} & \textbf{DSFL (Ours)} \\
                     & \cite{paillier1999}    & \cite{bonawitz2017}    & \\
\midrule
\textbf{Complexity}  & $O(N\cdot d)$          & $O(N^2)$               & $\bm{O(N \cdot m)}$ \\
\addlinespace
\textbf{Dropouts}    & Robust                 & Heavy recovery         & \textbf{Immediate (intra-shard)} \\
\addlinespace
\textbf{Poisoning}   & No defense             & No defense             & \textbf{Consistency Check} \\
\addlinespace
\textbf{Topology}    & Star                   & Full mesh              & \textbf{Sharded} \\
\addlinespace
\textbf{Comm. Cost}  & High (2048-bit)        & High (key vol.)        & \textbf{Low (1:1 shard)} \\
\bottomrule
\end{tabular}
\end{table}

\subsection{Audit: Financial Utility and Robustness}
The final aggregated global model was evaluated on a held-out test set comprising 20\% of
transactions unseen during training. This emulates a regulatory audit of the fraud detection
system. All reported metrics are mean values over five independent trials; standard deviations
are reported in parentheses.
\begin{itemize}
    \item \textbf{Global Recall:} 91.2\% ($\pm$0.8\%). The collaborative model significantly
    outperformed the local-model average of 68\% ($\pm$3.1\%), demonstrating successful
    cross-shard knowledge transfer.
    \item \textbf{Simulated Financial Impact:} In simulation on the Kaggle dataset,
    the global DSFL model correctly flagged transactions totalling \$120,450 that the
    local-only baseline would have approved. This figure is derived from the held-out test
    partition and should be interpreted as a relative performance indicator, not a
    real-world financial forecast.
    \item \textbf{Resilience to Failure:} Under a hostile 20\% dropout regime, the Active
    Recovery mechanism reconstructed all orphaned masks in under 0.8\,s. Final model utility
    (AUPRC) was statistically indistinguishable from the zero-dropout baseline
    ($p > 0.05$, paired $t$-test), confirming robustness without utility loss.
\end{itemize}

\begin{figure}[!t]
\centering
\includegraphics[width=\columnwidth]{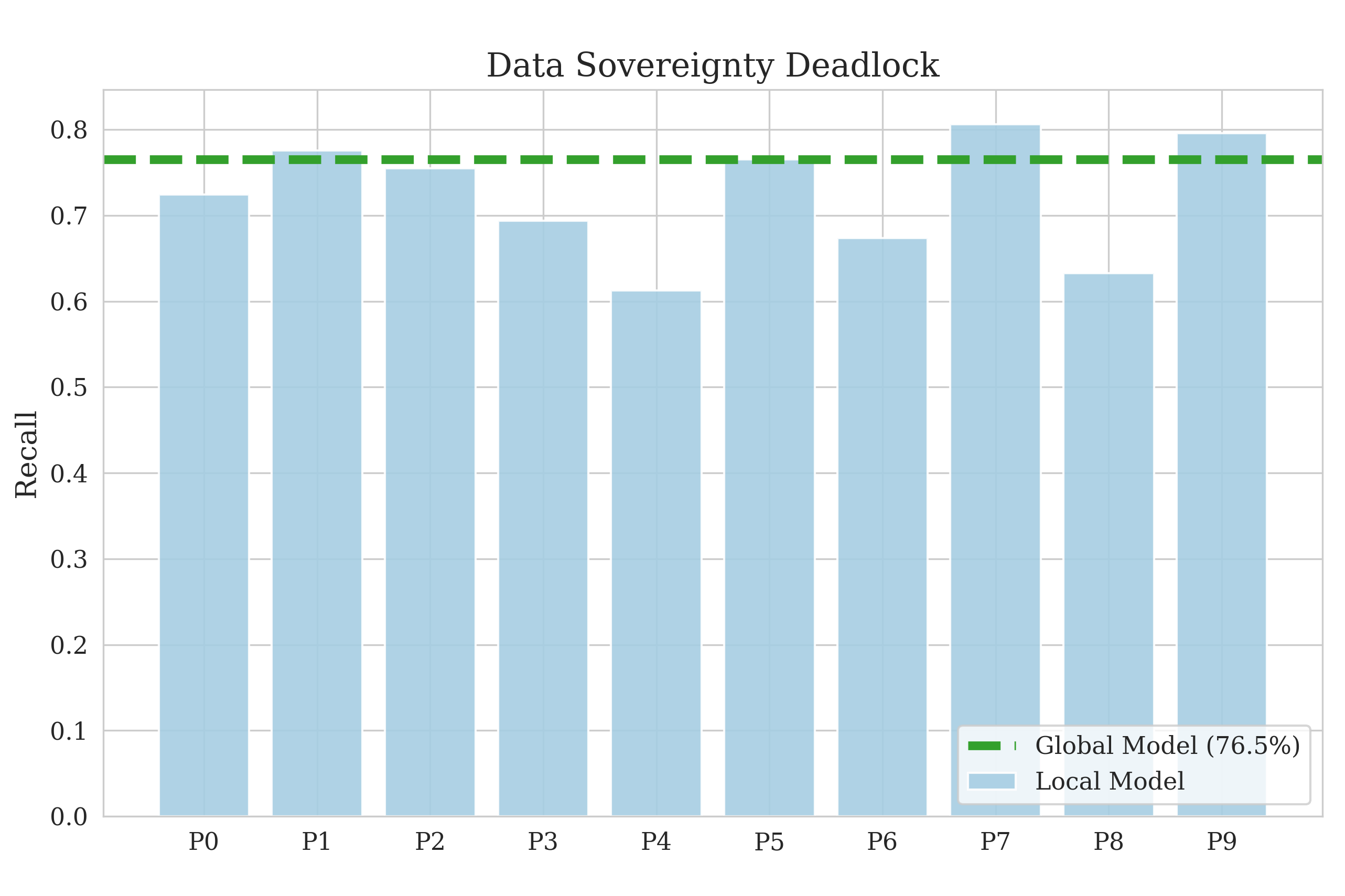}
\caption{\textbf{The Data Sovereignty Deadlock.} Recall of isolated local models (Blue) vs.\
the global DSFL model (Green) across ten banking partitions of the Kaggle dataset. Local models
exhibit high variance and poor detection of rare fraud patterns; the global model stabilizes
performance across the consortium.}
\label{fig:deadlock}
\end{figure}


\FloatBarrier 
\section{Conclusion and Future Work}

The central argument of this paper is that the scalability and
verifiability problems in federated fraud detection are not independent
and should not be solved separately. Existing approaches treat them
as such: homomorphic encryption schemes provide strong privacy but
are computation-bound; pairwise masking protocols achieve lighter
cryptography but require quadratic communication; and neither class
offers any mechanism to detect inconsistent or tampered updates.
DSFL's contribution is to resolve all three simultaneously through
a single architectural primitive—stochastic sharding—combined with
additive-homomorphic linear integrity tags that piggyback on the
masking structure at negligible additional cost.
 
We are careful to bound our claims accurately. The Linear Integrity
Tags detect \emph{consistency violations}: a submitted update that
does not match its declared tag will be rejected with probability
$1 - \mathrm{negl}(\lambda)$. They do not detect a Byzantine client
who submits a semantically harmful but arithmetically consistent
gradient. Addressing that class of attack requires robust aggregation
rules applied to the decrypted aggregate—a natural extension of
this work, discussed below.

The future work worth pursuing: (1) the more immediate is
tighter integration with hardware-based Trusted Execution
Environments. Intel~SGX, for instance, could protect the ECDH
key-exchange phase against a compromised host OS—a threat that
the current protocol does not address, since it assumes a secure
channel at the TLS layer but not below it. (2) the current protocol synchronizes
on the slowest shard in each round, which creates a predictable
latency floor in heterogeneous banking networks. An asynchronous
variant that allows faster shards to contribute partial aggregates
while slower ones catch up would be particularly valuable in
high-frequency payment clearing contexts, where round latency
directly determines how quickly fraud pattern updates propagate
to all participating institutions.

\bibliographystyle{IEEEtran}
\bibliography{references}

@InProceedings{bonawitz2017,
author = {Bonawitz, Keith and Ivanov, Vladimir and Kreuter, Ben and Marcedone, Antonio and McMahan, H. Brendan and Patel, Sarvar and Ramage, Daniel and Segal, Aaron and Seth, Karn},
title = {Practical Secure Aggregation for Privacy-Preserving Machine Learning},
year = {2017},
isbn = {9781450349468},
publisher = {Association for Computing Machinery},
address = {New York, NY, USA},
url = {https://doi.org/10.1145/3133956.3133982},
doi = {10.1145/3133956.3133982},
booktitle = {Proceedings of the 2017 ACM SIGSAC Conference on Computer and Communications Security},
pages = {1175–1191},
numpages = {17},
location = {Dallas, Texas, USA},
series = {CCS '17}
}

@inbook{zhu2019,
author = {Zhu, Ligeng and Liu, Zhijian and Han, Song},
title = {Deep leakage from gradients},
year = {2019},
publisher = {Curran Associates Inc.},
address = {Red Hook, NY, USA},
booktitle = {Proceedings of the 33rd International Conference on Neural Information Processing Systems},
articleno = {1323},
numpages = {11}
}

@InProceedings{geiping2020,
author = {Geiping, Jonas and Bauermeister, Hartmut and Dr\"{o}ge, Hannah and Moeller, Michael},
title = {Inverting gradients - how easy is it to break privacy in federated learning?},
year = {2020},
isbn = {9781713829546},
publisher = {Curran Associates Inc.},
address = {Red Hook, NY, USA},
booktitle = {Proceedings of the 34th International Conference on Neural Information Processing Systems},
articleno = {1421},
numpages = {11},
location = {Vancouver, BC, Canada},
series = {NIPS '20}
}

@inproceedings{bagdasaryan2020,
  title     = {How To Backdoor Federated Learning},
  author    = {Bagdasaryan, Eugene and Veit, Andreas and Hua, Yiqing and Estrin, Deborah and Shmatikov, Vitaly},
  booktitle = {Proceedings of the Twenty Third International Conference on Artificial Intelligence and Statistics},
  pages     = {2938--2948},
  year      = {2020},
  volume    = {108},
  series    = {Proceedings of Machine Learning Research},
  publisher = {PMLR},
  url       = {https://proceedings.mlr.press/v108/bagdasaryan20a.html}
}

@inproceedings{paillier1999,
author="Paillier, Pascal",
editor="Stern, Jacques",
title="Public-Key Cryptosystems Based on Composite Degree Residuosity Classes",
booktitle="Advances in Cryptology --- EUROCRYPT '99",
year="1999",
publisher="Springer Berlin Heidelberg",
address="Berlin, Heidelberg",
pages="223--238",
isbn="978-3-540-48910-8"
}

@article{shamir1979,
author = {Shamir, Adi},
title = {How to share a secret},
year = {1979},
issue_date = {Nov. 1979},
publisher = {Association for Computing Machinery},
address = {New York, NY, USA},
volume = {22},
number = {11},
issn = {0001-0782},
url = {https://doi.org/10.1145/359168.359176},
doi = {10.1145/359168.359176},
journal = {Commun. ACM},
month = nov,
pages = {612–613},
numpages = {2}
}

@misc{ulb2018fraud,
  author       = {{Machine Learning Group - ULB} and Yann-A{\"e}l Le Borgne and Andrea Dal Pozzolo and Olivier Caelen and Gianluca Bontempi},
  title        = {Credit Card Fraud Detection Dataset},
  year         = {2018},
  howpublished = {\url{https://www.kaggle.com/datasets/mlg-ulb/creditcardfraud}},
  note         = {Transactions made by European cardholders in September 2013; 284,807 transactions with 492 fraud cases}
}
\end{document}